\documentclass[preprint,aps] {revtex4}
\usepackage{amsmath}
\usepackage{graphicx}

\begin {document}

\title{COMPLETE GLAUBER CALCULATIONS OF REACTION AND INTERACTION CROSS SECTIONS IN LIGHT ION COLLISIONS}
\author{I. S. Novikov}
\email{ivan.novikov@wku.edu}
\affiliation{Department of Physics and Astronomy, Western Kentucky University, \\
1906 College Heights Blvd, \#11077, Bowling Green, KY 42101-1077, USA}
\author{Yu. Shabelski}
\affiliation{Petersburg Nuclear Physics Institute, NCR Kurchatov Institute \\
Gatchina, St.Petersburg 188350, Russia}

\begin{abstract}
We calculate the differences between reaction and interaction cross sections in the collisions of relativistic light ions with $\text{A}<40$ in the framework of Glauber theory. Although, in the optical approximation of Glauber theory these differences are approximately 1\% of the reaction cross sections, they increase up to 3-4\% when all scattering diagrams of Glauber theory are included in calculation. 
\end{abstract}


\maketitle

\section{\label{sec:Intro}Introduction}

Various nuclear targets were used  to measure interaction cross-sections with stable and unstable light isotopes~\cite{Oza,Oza1}. The obtained data was used to determine parameters of the nuclear density distribution. To do so, the Glauber theory was often used, see \cite{Gl1,Sit1,Gl2}.  The reaction cross sections were calculated in the framework of the Glauber theory and compared with the experimental data on the interaction cross sections. The values of parameters for nuclear matter density can be found in~\cite{CLS,ABV,DeDe}.

The conducted in~\cite{Oza,Oza1,MNS} analysis relied on the fact that the difference between interaction and reaction cross section is negligible.  Previously, the difference between $\sigma^{(I)}_{AB}$ and $\sigma^{(R)}_{AB}$ has been estimated to be less than a few percent, see~\cite{KSh,Oga}.  In \cite{KIO} authors analyzed the reaction and interaction cross section using the "black-sphere" model and rectangular uniform distribution of nuclear matter for nuclei with A $<$ 80.  It was shown that the difference in cross sections is approximately 60 mb, or 4-6\% of the reaction cross section $\sigma^{(R)}_{AB}$. 

Since the "black-sphere" model does not proved adequate framework for the calculation of reaction and interaction cross sections, we compute the difference between $\sigma^{(I)}_{AB}$ and $\sigma^{(R)}_{AB}$ using exact expressions of the Glauber theory as well as expressions obtained in the optical approximation.

\section{\label{sec:Section2}Elastic and inelastic nucleus-nucleus scattering in the Glauber theory}

In this section we provide expressions for the reaction, $\sigma^{(R)}_{AB}$, and interaction, $\sigma^{(I)}_{AB}$, cross sections and the difference between them obtained in the framework of the Glauber theory. Detailed derivation of these expressions can be found in \cite{ANS}.

\subsection{\label{subsec:Subsection2.1}Reaction cross section $\sigma^{(R)}_{AB}$ of light ion collisions}

In the Glauber theory, the reaction cross section can be derived using expression for the elastic scattering amplitude. As a result we obtain
\begin{equation}
\label{13}
\sigma^{(R)}_{AB} = \sigma^{tot}_{AB} - \sigma^{el}_{AB} = \int d^2b [1 - \vert S_{AB}(b)\vert^2],
\end{equation}
where
\begin{equation}
\label{eq2}
S_{AB}(b) = \langle A \vert \langle B \vert \left\{\prod_{i, j} [1 - \Gamma_{NN}(b + u_i - s_j)] \right\} \; \vert B \rangle \vert A \rangle
\end{equation}
with
\begin{equation}
\label{eq3}
\Gamma_{NN}(b + u_i - s_j) = \frac{1}{2 \pi ik} \int d^2q \hspace{0.1cm} e^{-iq(b + u_i - s_j)} f^{el}_{NN}(q),
\end{equation}
$f^{el}_{NN} (q)$ is the amplitude of elastic nucleon-nucleon scattering and $u_i$ and $s_j$ are the transverse coordinates of nucleons.

In the discussion presented below we used the standard assumptions of the Glauber theory (see for example in \cite{CzMa}). First of all,  we consider the phase of A-B scattering to be equal to the sum of the phases for the scattering of the nucleons of the A-nucleus on the nucleons of the B-nucleus. Secondly, we assume that the 3-dimensional distribution of nucleons in projectile and target nuclei can be expressed as a product of normalized 3-dimensional single-particle densities $\rho^{\prime}_A(u_i, u_{i,z})$ and $\rho^{\prime}_B(s_j, s_{j,z})$, where $u_{i,z}, s_{j,z}$ are z-components of the nucleon coordinates
\begin{eqnarray}
\label{eq5}
\rho^{\prime}_A(u_1, u_{1,z},\ldots,u_A, u_{A,z}) &=&\prod^A_{i=1} \rho^{\prime}_A(u_i, u_{i,z}); \nonumber\\
\rho^{\prime}_B(s_1, s_{1,z} \ldots,s_B, s_{B,z}) &=&\prod^B_{j=1} \rho^{\prime}_B(s_j, s_{j,z}); \\
\; \int d^3r_i \rho^{\prime}(r_i) &=&1.\nonumber
\end{eqnarray}
Since we consider forward scattering, we can integrate over longitudinal coordinates $u_{i,z}$ and $s_{j,z}$. In this case the expression for $S_{AB}$ as a function of impact parameter $b$ can be written in the following form
\begin{equation}
\label{SAB_int}
S_{AB}(b) = \idotsint \prod^A_{i=1} \rho_A(u_i) \left\{\prod_{i, j} [1 - \Gamma_{NN}(b + u_i - s_j)] \right\} \prod^B_{j=1} \rho_B(s_j) \prod_{i=1}^A du_i \prod_{j=1}^B ds_j,
\end{equation}
where $\rho_A\left( u_i \right)$ and $\rho_B\left( s_j\right)$ are 2-dimensional single-particle distributions of nucleons in projectile and target nuclei. 

In optical approximation, the integral form of $S_{AB}$ can be simplified:
\begin{equation}
\label{eq7}
S_{AB}^{opt}(b)  \approx \exp [- T_{opt}(b)],
\end{equation}
where
\begin{equation}
\label{eq8}
T_{opt}(b) = \frac{\sigma_{NN}^{tot}}{4 \pi \beta} \int d^2b_1 d^2b_2 T_A(b_1) T_B(b_2)\exp\left[-\frac{(b+b_1-b_2)^2}{2\beta}\right],
\end{equation}
with 
\begin{equation}
\label{eq9}
T_A(b) = A \int^{\infty}_{-\infty} dz \rho_A\left(\sqrt{b^2+z^2} \right).
\end{equation}

\subsection{\label{subsec:Section2.2}Interaction cross section $\sigma^{(I)}_{AB}$ of light ion collisions} 

The interaction cross sections is defined as a cross section of all processes which do not include target B-nucleus excitation or disintegration (these states are denoted as $B^*$). Therefore, the difference between interaction and reaction cross sections can be written as
\begin{equation}
\sigma^{(R)}_{AB} - \sigma^{(I)}_{AB} = \sigma(AB \to AB^*)
\end{equation}

Using framework of the Glauber theory, one can obtain following expressions for the $\sigma(AB \to AB^*)$:
\begin{equation}
\sigma_{AB \to AB^*} = \int [J_{AB}(b) - S^2_{AB}(b)] d^2b.
\end{equation}
where
\begin{eqnarray}
J_{AB}(b) = && \langle A \vert  \langle B \vert \left\{ \prod_{i,j} [1 - \Gamma_{NN} (b + u_i - s_j)] \right\}\vert A \rangle \times \nonumber\\
&& \times \langle A \vert \left\{ \prod_{i,j'} [1 - \Gamma_{NN} (b + u_i - s_j')] \right\} \vert B \rangle \vert A \rangle.
\end{eqnarray}

$J_{AB}$ can be simplified using optical approximation
\begin{equation}
J^{opt}_{AB} \approx  \exp\left(-T^{*}_{opt}(b) \right)\,,
\end{equation}
where 
\begin{eqnarray}
T^{*}_{opt}(b) = \frac{\sigma^{tot}_{NN}}{2\pi\beta} && \int d^2b_1 d^2b_2 T_A(b_1) T_B(b_2) e^{-\frac{(b+b_1-b_2)^2}{2\beta}} \times \nonumber\\
&& \times \left(1 - 2 \frac{\sigma^{el}_{NN}}{\sigma^{tot}_{NN}} \frac1B \int d^2b_3 T_B(b_3) e^{-\frac{(b+b_1-b_3)^2}{2\beta}}\right).
\end{eqnarray}

We also provide expression for the cross section of the $ \text{AB} \to \text{A}^* \text{B}^*$ processes. It can be obtained in the same manner as one for $\sigma(AB \to AB^*)$:
\begin{equation}
\sigma_{AB \to A^*B^*} = \int [I_{AB}(b) - S^2_{AB}(b)] d^2b.
\end{equation}
where
\begin{equation}
I_{AB}(b) = \langle A \vert  \langle B \vert \left\{ \prod_{i,j} [1 - \Gamma_{NN} (b + u_i - s_j)] \right\}^2 \vert B \rangle \vert A \rangle.
\end{equation}

In the optical approximation it reads:
\begin{equation}
I^{opt}_{AB}(b) \approx  \exp \left(-T^{**}_{opt}(b)\right),
\end{equation}
where
\begin{equation}
T^{**}_{opt}(b) = \frac1{2\pi\beta} \int d^2b_1 d^2b_2 T_A(b_1) T_B(b_2) \left(\sigma_{NN}^{tot} e^{-\frac{(b+b_1-b_2)^2}{2\beta}} - 2 \sigma^{el}_{NN} e^{-\frac{(b+b_1-b_2)^2}{\beta}}\right).
\end{equation}

\section{\label{sec:Section3}Numerical calculations of $\sigma^{(R)}$ and $\sigma^{(I)}$ difference}

Use of Monte Carlo (MC) integration technique for numerical calculation of \ref{SAB_int} was first proposed in~\cite{ZUS,Shm}. MC integration was used for cross section calculations in~\cite{Gar,AlLo,Sh11,Var}.

To evaluate $S_{AB}$, the set of randomly generated nucleon coordinates is required. The standard MC approach calls for set of coordinates uniformly distributed in the interaction region. That leads to a significant loss of accuracy and an increase in computational time due to the fact that often several nucleon coordinates give negligible contribution to the $S_{AB}$.

The use of Monte Carlo Markov Chain (MCMC) algorithm for generation of nucleon coordinates $u_i$ and $s_j$ distributed according to density distributions $\rho_A(u_i)$ and $\rho_B(s_j)$ proved to be a more efficient way to calculate $S_{AB}$. The Metropolis-Hastings algorithm~\cite{Metro} was used to obtain a sequence of random numbers from a defined distributions $\rho_A$ and $\rho_B$. Generated set of coordinates were used to calculate average value of  $S_{AB}$. We apply the same technique to $\sigma_{AB \to A^*B^*}$ and $\sigma_{AB \to AB^*}$ calculations.

It is common to use the following parametrization for the amplitude of the elastic nucleon-nucleon scattering $f^{el}_{NN} (q)$
\begin{equation}
\label{eq4}
f^{el}_{NN} (q) = \frac{ik\sigma^{tot}_{NN}}{4 \pi} \exp \left( -\frac12 \beta q^2\right),
\end{equation}
where $\sigma^{tot}_{NN}$ is the total nucleon-nucleon cross section, and $\beta$ is the slope parameter of the differential nucleon-nucleon cross-section dependence on $q^2$.  We neglect the real part of $f^{el}_{NN} (q)$ since it gives negligible contribution to the reaction cross section. The parameters of NN elastic scattering amplitude Eq.~(\ref{eq4}) at energy about 1 GeV were taken as 
\begin{equation}
\sigma^{tot}_{NN} = 43 \text{ mb} \;,\; \sigma^{el}_{NN} = 24.3 \text{ mb} \;,\; \beta = 0.35 \text{ fm}^2
\end{equation}

The results of the nuclear cross section calculations depend upon the shape of the nuclear density distribution, see \cite{MNS}. All calculations presented below were done with Woods-Saxon distribution, \cite{WoSa}, for projectile and target nuclei:
$$
\rho_A (r) = \frac{\rho_0}{1+\exp\left((r-c)/a\right)}.
$$
In all calculations the value of the parameter $a$ was set to be equal 0.54 fm.

The $c$-parameter of  Woods-Saxon distribution was varied to match calculated value for the reaction cross section $\sigma^{(R)}_{AB}$,  with the experimental value for interaction cross section $\sigma^{(I)}_{\text{exp}}$ presented in~\cite{Oza,Oza1}. Reaction cross section $\sigma^{(R)}_{AB}$ was calculated using the expression for $S_{AB}$ obtained in optical approximation Eq.~\ref{eq7} and complete expression of the Glauber theory (Eq.~\ref{eq2}). Experimental cross section, $\sigma_{\text{exp}}^{(I)}$, as well as the values of parameter $c$ calculated in optical approximation (OA) and using complete expressions of Glauber theory, are presented in Table \ref{tab:table1}. Obtained results for parameters of Woods-Saxon distribution are in agreement with the values for $R_{rms}$ provided in \cite{MNS}.

\begin{table}[h!]
\caption{\label{tab:table1}Experimental interaction cross section $\sigma^{(I)}_{\text{exp}}$ (\cite{Oza,Oza1}) and calculated in optical approximation (OA) and in complete Glauber Theory values of Woods-Saxon parameter $c$ (in fm).}
{\begin{tabular}{@{}cccc@{}} \toprule
Nucleus &  $\sigma^{(I)}_{\text{exp}}$, mb & $c_{\text{OA}}$, fm & $c_{\text{Gl}}$, fm \\
\colrule
C$^{12}$ & $853 \pm 6$ & 0.85 & 1.39 \\ 
N$^{14}$ & $932 \pm 9$ & 1.40 & 1.75 \\
O$^{16}$ & $982 \pm 6$ & 1.50 & 1.90 \\
F$^{19}$ & $1043 \pm 24$ & 1.70 & 2.10 \\
Mg$^{24}$ & $1136 \pm 72$ & 1.88 & 2.25\\
Cl$^{35}$ & $1327 \pm 14$  & 2.43 & 2.75\\
\botrule
\end{tabular}}
\end{table}

Calculated parameters ($c_{\text{OA}}$ and $c_{\text{Gl}}$) were used to calculate $\sigma_{AB \to AB^*}$ and $\sigma_{AB \to A^*B^*}$ cross sections for the corresponding stable isotopes. Calculated cross sections $\sigma_{AB \to A^*B^*}$ and $\sigma_{AB \to AB^*}$ and computational errors are presented in Table~\ref{tab:table2}.

\begin{table}[h!]
\caption{\label{tab:table2}Calculated in optical approximation (OA) and in complete Glauber Theory $\sigma_{AB \to AB^*}$ and $\sigma_{AB \to A^*B^*}$ cross sections (in mb)}
{\begin{tabular}{@{}ccccc@{}} \toprule
Nucleus & \multicolumn{2}{c}{Optical Approximation} & \multicolumn{2}{c}{Glauber Theory} \\ \cline{2-5}
 & $\sigma_{AB \to A^*B^*}$  & $\sigma_{AB \to AB^*}$ & $\sigma_{AB \to A^*B^*}$ & $\sigma_{AB \to AB^*}$ \\   
\colrule
C$^{12}$  & $169 \pm 2$ & $10.4 \pm 1.0$ &  $102 \pm 1$ & $34 \pm 5$ \\ 
N$^{14}$  & $180 \pm 2$ & $7.7 \pm 1.0$ &  $107 \pm 2$ &  $34 \pm 6$ \\ 
O$^{16}$  & $185 \pm 2$ & $7.9 \pm 1.0$ & $110 \pm 2$ &  $33 \pm 4$\\ 
F$^{19}$  & $188 \pm 3$ & $6.9 \pm 1.3$ &  $115 \pm 3$ &  $36 \pm 3$ \\ 
Mg$^{24}$  & $192 \pm 2$ & $6.8 \pm 1.4$ & $118 \pm 1$ &  $41 \pm 6$\\ 
Cl$^{35}$  & $208 \pm 4$ & $5.2 \pm 1.0$ & $128 \pm 2$ &  $42 \pm 5$\\ 
\botrule
\end{tabular}}
\end{table}

The difference between the values for reaction cross section $\sigma^{(R)}_{AB}$ obtained in optical approximation and using complete expression of Glauber theory differ by approximately 10\%, see in~\cite{MNS}. Cross sections of the $AB \to AB^*$  process also depend on the used approximation. However, values of  $\sigma_{AB \to AB^*}$ obtained by using the expression of the Glauber theory are significantly higher than cross sections calculated in optical approximation.

Since in $AB \to AB^*$ collisions the projectile nucleus stays in its ground state, we conclude that these processes are peripheral. In other words, collisions with an impact parameter approximately equal to the sum of the projectile and target nucleus radii contribute more to the $\sigma_{AB \to AB^*}$ cross section than collisions with small impact parameters. To illustrate this fact, the contribution of different impact parameters to the reaction cross section $\sigma^{(R)}_{AB}$, $\sigma_{AB \to AB^*}$ and $\sigma_{AB \to A^*B^*}$ are shown in Figure~1. All contributions to cross sections were calculated for the  $^{12}$C-$^{12}$C in optical approximation. The error of the MCMC calculation for the  $\sigma_{AB \to AB^*}$ cross section is shown on the graph.
%Figure2
\begin{figure}[h!]
\label{pic:fig2}
\includegraphics[width=0.6\textwidth]{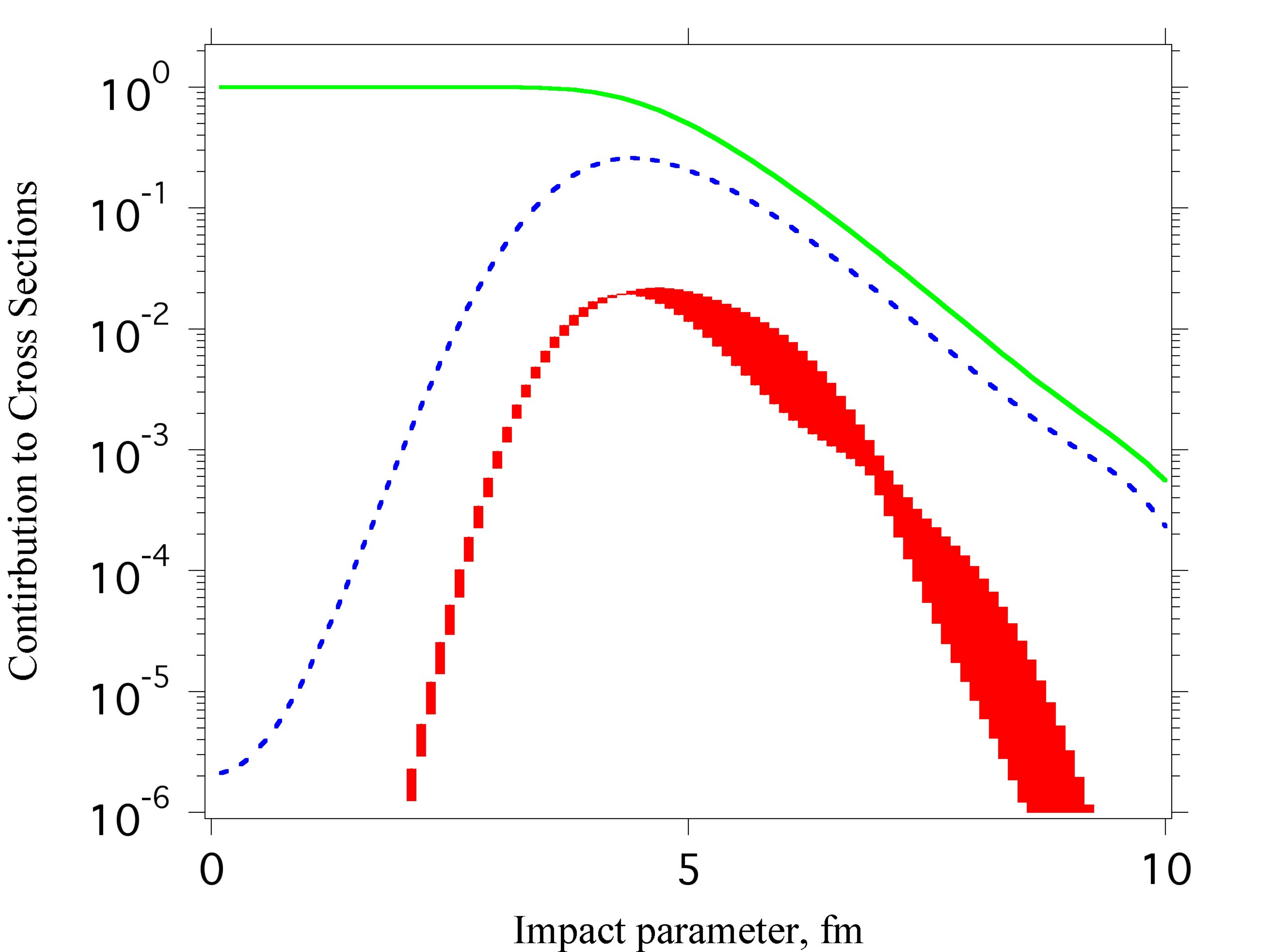}
\caption{Contribution of different impact parameters to the total reaction cross section (solid green), to the $\sigma_{AB \to A^*B^*}$ cross section (dashed blue) and to the $\sigma_{AB \to AB^*}$ cross section (red bars) for $^{12}$C-$^{12}$C interaction (in semi-log scale).}
\end{figure}

Obtained results are in agreement with the ones presented in \cite{KSh}, where contribution of impact parameters to $\sigma^{(R)}_{AB}$ and  $\sigma_{AB \to AB^*}$ cross sections were calculated for $^{16}$O-Al and $^{16}$O-Pb reactions at 200 GeV per nucleon.

For the peripheral process we expect the following dependence of $\sigma_{AB \to AB^*}$ on the atomic weight of the projectile nucleus:
$$
\sigma_{AB \to AB^*} \sim  {\rm R_A} \sim {\rm A}^{1/3},
$$
when $A \gg 1$. Therefore, the ratio between $\sigma_{AB \to AB^*}$ and $\sigma_{AB}^{(R)}$ is expected to depend on atomic weight as
\begin{equation}
R^*(A) = \sigma_{AB \to AB^*}/\sigma_{AB}^{(R)} \sim {\rm A}^{-n} 
\end{equation}
with $n \sim  1/3$. The A-dependence of ratio $R^*(A)$ calculated in optical approximation and using complete expressions of the Glauber theory are presented in Figure~2.  

%Figure3
\begin{figure}[pt]
\includegraphics[width=0.6\textwidth]{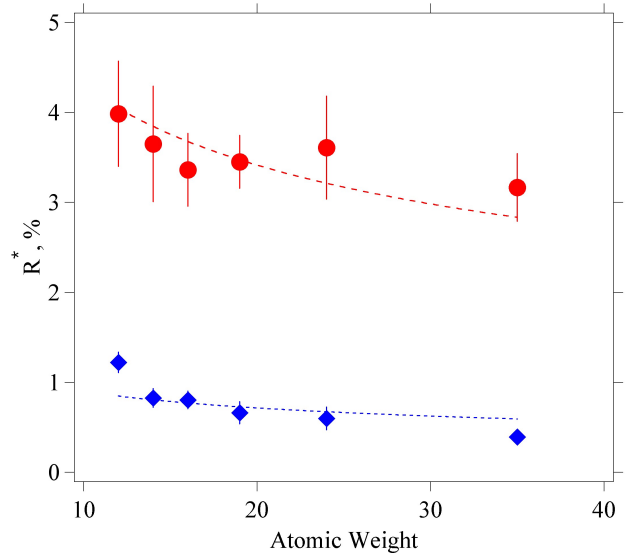}
\caption{The values of the $R^*$ calculated in optical approximation (blue diamonds) and in complete Glauber theory (red circles) as the functions of projectile atomic weight. Data fitting with $\alpha A^{-n}$ function, where $n = 1/3$, is shown as dashed line.}
\end{figure}

\section{\label{sec:Conclusion}Conclusion}

The differences of $\sigma_{AB}^{(R)}$ and $\sigma_{AB}^{(I)}$ were calculated for light (A = 12-35) stable isotopes interacting with C$^{12}$ target at approximately 1000 MeV per nucleon. The calculations were done in the optical approximation and using complete expressions of the Glauber theory. 

The cross sections $\sigma_{AB \to AB^*}$ obtained in optical approximation are approximately 1$\%$ of the reaction cross section. It is in agreement with the results presented in \cite{Oga}. However,  the values calculated using complete expressions of the Glauber theory are  approximately 3-4$\%$ of the reaction cross section. The absolute values of the $\sigma_{AB \to AB^*}$ calculated using complete expressions of the Glauber theory are approximately 30$-$40 mb. 

When the Glauber theory is used to analyze experimental data on interaction cross section, it is necessary to correct the value of the calculated reaction cross section as
\begin{equation}
\sigma_{AB}^{(R)} \geq \sigma_{AB}^{(I)} + \sigma_{AB \to AB^*}.
\end{equation}
This leads to the increase of the projectile the $R_{\text{rms}}$ radii by approximately 2\% or 0.05 fm for nuclei with atomic weight A $<$ 40. This difference can be additionally increased if the projectile nucleus has exited states \cite{Oza1}.

Since  $AB \to AB^*$  processes are very peripheral, it is possible that the difference between interaction and reaction cross sections increases in the nuclei with halo or skin structures. Hence, we believe that when the parameters of nuclear halo and skin are extracted from the experimental data, the difference between interaction and reaction cross sections should not be neglected.

This work was supported in part by grant RFBR 11-02-00120-a.

\end{document}